\documentclass[amsmath,amssymb,twocolumn,superscriptaddress,prb]{revtex4-1}%,longbibliography
\usepackage{physics}
\usepackage{bm}
\usepackage{graphicx}
\usepackage{xcolor}
\usepackage{verbatim}
\usepackage{MnSymbol}
\usepackage{diagbox}
\usepackage[colorlinks=true, breaklinks=true, linkcolor=red, citecolor=blue, urlcolor=blue]{hyperref}

\begin{document}
\title{Real-frequency TPSC+DMFT investigation of the square-lattice Hubbard model}
\author{Lei Geng}
\affiliation{Department of Physics, University of Fribourg, 1700 Fribourg, Switzerland}
\author{Jiawei Yan}
\affiliation{2020 X-Lab, Shanghai Institute of Microsystem and Information Technology, Chinese Academy of Sciences, Shanghai 200050, China}
\author{Philipp Werner}
\affiliation{Department of Physics, University of Fribourg, 1700 Fribourg, Switzerland}

\begin{abstract}
We investigate the two-dimensional Hubbard model using a real-frequency implementation of the TPSC+DMFT approach. This hybrid method combines the nonlocal correlations captured by the Two-Particle Self-Consistent (TPSC) approach with the local dynamical correlations of Dynamical Mean-Field Theory (DMFT). The results demonstrate that TPSC+DMFT effectively describes pseudogap physics and nonlocal fluctuations in the moderately correlated regime, while also reproducing the Mott insulating state at larger interaction strengths. For doped Mott insulators, we find that TPSC+DMFT captures the evolution of Fermi pockets into Fermi arcs, consistent with the results from cluster DMFT and photoemission studies. These findings highlight the capability of TPSC+DMFT to bridge the gap between weak and strong coupling physics in Hubbard models, providing insights into spin and charge fluctuations, as well as their role in the pseudogap formation.
\end{abstract}
\maketitle

\section{Introduction}

The Hubbard model is one of the cornerstone models for the study of strongly correlated electron systems. Despite its deceptively simple mathematical form, the model captures a wide range of emergent phenomena, including superconductivity, magnetism, and Mott insulating behavior with appropriate parameters~\cite{baeriswyl1995}. Consequently, finding accurate solutions to the Hubbard model has remained an important problem in theoretical condensed matter physics. In one dimension, the exact solution for the ground state of the Hubbard model was provided by Lieb and Wu using the Bethe Ansatz~\cite{lieb1968absence}, and extensions to few-leg ladders have been studied using the density matrix renormalization group method \cite{noack1996}. In the limit of infinite dimensions, dynamical mean-field theory (DMFT) has been established as an exact numerical framework~\cite{metzner1989correlated,georges1996dynamical}. However, for more realistic two- and three-dimensional systems, obtaining accurate solutions and clarifying the physics of the Hubbard model remains a formidable challenge~\cite{Schaefer2021}. 

Over the years, various numerical approaches have been developed, including diagrammatic methods \cite{vanhoucke2010}, tensor network methods \cite{schollwoeck2005}, quantum Monte Carlo simulations \cite{gubernatis2016}, and various extensions of DMFT \cite{maier2005,rohringer2018}. These methods can sometimes be synergistically combined to exploit their respective strengths. For instance, DMFT captures local correlations, but ignores nonlocal spatial correlations. A class of hybrid approaches integrates spatial correlations from methods like GW \cite{hedin1965} or the fluctuation exchange (FLEX) approximation~\cite{bickers1989conserving} with the local dynamical correlations captured by DMFT, showing promising results under specific conditions~\cite{boehnke2016strong,biermann2003first,petocchi2020normal,kitatani2015flex+,gukelberger2015dangers}. In this work, we focus on the combination of the two-particle self-consistent theory (TPSC)~\cite{vilk1997non,yan2024spin} with DMFT. The TPSC approach has been successfully used to study the moderately correlated two-dimensional Hubbard model. It employs renormalized spin and charge vertices, satisfied sum rules and the Mermin-Wagner theorem, and captures the pseudogap phenomenon known from cuprate physics~\cite{kyung2004pseudogap}. TPSC itself is however only applicable to systems with weak or moderate correlation strength. To extend it deeper into the correlated regime, TPSC has recently been combined with DMFT~\cite{martin2023nonlocal,simard2023dynamical,zantout2023improved}, either by employing the DMFT double occupation in the TPSC sum rules, or by additionally replacing the local self-energy component by the DMFT result. 

Here, we explore different variants of the TPSC+DMFT approach using a real-frequency implementation which provides direct access to spectral functions. We find that a combination at the ``one-shot" level provides a meaningful description of the half-filled moderately correlated system and of the doped Mott insulator in the optimally to overdoped regimes. In particular, this method captures the momentum dependent life-times of quasi-particles and the Fermi arc phenomenon. Self-consistent variants such as TPSC+GG~\cite{simard2022nonequilibrium,Schaefer2021,yan2024spin} and the corresponding TPSC+GG+DMFT scheme underestimate nonlocal correlations and miss the pseudogap physics. 

The rest of the paper is organized as follows. In Sec.~\ref{section2}, we present the Hubbard model and various numerical methods, including DMFT, TPSC, TPSC+DMFT and FLEX. In Sec.~\ref{section3A} and Sec.~\ref{section3B}, we use these methods to investigate half-filled systems, covering a range of interaction strengths from moderate to strong. In Sec.~\ref{section3C}, we shift our attention to a cuprate-inspired doped Hubbard model with next-nearest-neighbor hopping. In Sec.~\ref{section4}, we present a brief conclusion, while additional results from self-consistent variants of TPSC and TPSC+DMFT are provided in the Appendix.

\section{Model and Methods}\label{section2}

We begin by introducing the two-dimensional Hubbard model and its relevant properties, followed by a detailed description of the methods employed in this study: DMFT, TPSC, TPSC+DMFT, and FLEX.  The implementation of these methods on the real-frequency axis is also discussed. 

\subsection{The Hubbard model}

In real space, the single-band Hubbard model is described by the Hamiltonian 
\begin{equation}
H=\sum_{i,j,\sigma}t_{i,j}(c_{i,\sigma}^{\dagger}c_{j,\sigma}+\text{h.c.})+U\sum_{i}n_{i,\uparrow}n_{i,\downarrow}-\mu\sum_{i,\sigma} n_{i,\sigma}.
\label{Eq1}
\end{equation}
Here, the $c_{i,\sigma}$ ($c_{i,\sigma}^{\dagger}$) are the annihilation (creation) operators for electrons with spin $\sigma$, and $n_{i,\sigma}$ represents the density operators for the spin-$\sigma$ electrons. $U$ denotes the on-site interaction strength, $\mu$ is the chemical potential, and $t_{i,j}$ is the hopping amplitude between the sites $j$ and $i$. In this work, we focus on the two-dimensional model, where each site has four nearest neighbors and four next-nearest neighbors. The corresponding hopping parameters are denoted by $t$ and $t'$, respectively. The first term in Eq.~\eqref{Eq1} is often expressed in momentum space as $\sum_{\mathbf{k},\sigma}\epsilon_{\mathbf{k}} c_{\mathbf{k},\sigma}^{\dagger}c_{\mathbf{k},\sigma}$, where
\begin{equation}
\epsilon_{\mathbf{k}}=-2t\left[\cos(k_x)+\cos(k_y)\right]-4t'\cos(k_x)\cos(k_y)
\label{Eq2}
\end{equation}
is the noninteracting dispersion. We use the nearest-neighbor hopping $t=1$ as the unit of energy throughout the paper and set the lattice spacing to unity.

In the two-dimensional Hubbard model, nonlocal correlations play a significant role, making DMFT less reliable than in the three-dimensional case. Hence, the two-dimensional model is a good testing ground for various extensions of DMFT. Due to the Mermin-Wagner theorem, antiferromagnetic or spin-density-wave transitions are forbidden at nonzero temperature in the two-dimensional Hubbard model. This is violated by the Hartree(-Fock) and DMFT descriptions. We thus stay above the corresponding N\'eel temperatures and impose the paramagnetic solution in DMFT throughout this study.
 
\subsection{DMFT}

DMFT neglects the spatial dependence of the self-energy, while retaining its temporal dependence, in contrast to static mean-field theories~\cite{georges1996dynamical}. This enables DMFT to describe a Mott-insulating state. DMFT is exact in the limit of infinite dimensions~\cite{metzner1989correlated} but provides only an approximate description of (interacting) two-dimensional systems. By assuming a spatially local self-energy~$\Sigma(\omega)$ \cite{mueller-hartmann1989}, the lattice model is mapped onto a single-site Anderson impurity model, where the hoppings between the impurity and the bath sites are described by a hybridization function, $\Delta(\omega)$. Solving this impurity problem requires an impurity solver to compute the impurity Green's function for a given $\Delta(\omega)$ and $U$. In this work, we use the non-crossing approximation (NCA)~\cite{li2021,eckstein2010nonequilibrium}, which is well-suited for systems with strong interactions. 

The impurity Green's function can be written in terms of the self-energy and hybridization function as 
\begin{equation}
    G_{\mathrm{imp}}(\omega)=\frac{1}{\omega+\mu-\Delta(\omega)-\Sigma_{\mathrm{imp}}(\omega)}.
    \label{Eq3}
\end{equation}
To complete the self-consistency loop, we also need to consider the (local) Green's function of the lattice model, 
\begin{equation}
\begin{split}
G(\mathbf{k},\omega)&=\frac{1}{\omega+\mu-\epsilon_{\mathbf{k}}-\Sigma_{\mathrm{loc}}(\omega)},\\
G_{\mathrm{loc}}(\omega)&=\frac{1}{N_{\mathbf{k}}}\sum_{\mathbf{k}}G(\mathbf{k},\omega),
\end{split}
\label{Eq4}
\end{equation}
where $N_{\mathbf{k}}$ is the number of $\mathbf{k}$ points.
The self-consistency condition for the local Green's function and the DMFT approximation for the self-energy can then be written as
\begin{equation}
\begin{split}
G_{\mathrm{loc}}(\omega)&=G_{\mathrm{imp}}(\omega),\\
\Sigma_{\mathrm{loc}}(\omega)&=\Sigma_{\mathrm{imp}}(\omega).
\end{split}
\label{Eq5}
\end{equation}

The self-consistency loop starts by solving the impurity problem for the given hybridization function (in the first loop, $\Delta(\omega)$ can be set to zero). Then, the atomic Green's function $G_{\mathrm{ato}}(\omega)$ is obtained from the impurity Green's function $G_{\mathrm{imp}}(\omega)$, which is subsequently used to compute the lattice Green's function $G(\mathbf{k},\omega)$ in Eq.~\eqref{Eq4} via the Dyson equation
\begin{equation}
\begin{split}
G_{\mathrm{ato}}^{-1}(\omega)&=G_{\mathrm{imp}}^{-1}(\omega)+\Delta(\omega),\\
G^{-1}(\mathbf{k},\omega)&=G_{\mathrm{ato}}^{-1}(\omega)-\epsilon_{\mathbf{k}}.
\end{split}
\label{Eq6}
\end{equation}
In this way, we avoid the explicit calculation of the self-energy, which may have significant numerical errors at high frequencies~\cite{labollita2023stabilizing}. Finally, the local Green's function $G_{\mathrm{loc}}(\omega)$ can be obtained from the momentum-resolved Green's function $G(\mathbf{k},\omega)$ by Eq.~\eqref{Eq4} and the hybridization function $\Delta(\omega)$ is updated using the local Green's function
\begin{equation}
\Delta(\omega)=G_{\mathrm{ato}}^{-1}(\omega)-G_{\mathrm{loc}}^{-1}(\omega).
\label{Eq7}
\end{equation}
This loop is iterated until the Green's function converges.

\subsection{TPSC}

The TPSC method can be viewed as an improved version of the random phase approximation (RPA)~\cite{wolff1960spin,bickers1989conserving} for calculating spin and charge susceptibilities. Unlike the RPA, the TPSC framework is designed to satisfy fundamental constraints, such as the Pauli exclusion principle and the Mermin-Wagner theorem, ensuring a more physically consistent description~\cite{vilk1997non}.

We define the spin operator $S_i^z$ and charge operator $n_i$ at site $i$ of our single orbital model as
\begin{equation}
    \begin{split}
    \hat{S}_i^z \equiv \hat{n}_{i,\uparrow}-\hat{n}_{i,\downarrow},\\
    \hat{n}_i \equiv \hat{n}_{i,\uparrow}+\hat{n}_{i,\downarrow}.
    \end{split}
    \label{Eq8}
\end{equation}
The corresponding spin and charge susceptibilities are $\chi^{\mathrm{sp}}$ and $\chi^{\mathrm{ch}}$~\cite{yan2024spin},
\begin{equation}
    \begin{split}
    \chi^{\mathrm{sp}}_{i,j}(t,t') &= -\mathrm{i}\left<\mathcal{T}\left[\hat{S}_i^z(t)\hat{S}_j^z(t')\right]\right>,\\
    \chi^{\mathrm{ch}}_{i,j}(t,t') &= -\mathrm{i}\left<\mathcal{T}\left[\hat{n}_i(t)\hat{n}_j(t')\right]\right>+\mathrm{i}n_i(t)n_j(t'),
    \end{split}
    \label{Eq9}
\end{equation}
where $\mathcal{T}$ represents the time-ordering operator on the Schwinger-Keldysh contour. Within the RPA framework, the spin and charge susceptibilities can be derived from the bare susceptibility $\chi^0$ as
\begin{equation}
    \begin{split}
    \chi^{\mathrm{sp}} &= \frac{\chi^0}{1-\frac{1}{2}U\chi^0},\\
    \chi^{\mathrm{ch}} &= \frac{\chi^0}{1+\frac{1}{2}U\chi^0}.
    \end{split}
    \label{Eq10}
\end{equation}
Here, the bare susceptibility is defined as $\chi^0_{i,j}(t,t')=-2\mathrm{i}G^0_{i,j}(t,t')G^0_{j,i}(t',t)$~\cite{bickers1989conserving}, where $G^0$ denotes the bare Green's function obtained from the Hartree-Fock approximation (the Fock term is absent in the Hubbard model).

It is straightforward to verify that the diagonal elements of the spin and charge susceptibilities, as expressed in Eq.~\eqref{Eq10}, do not satisfy the Pauli exclusion principle $n_{i,\sigma}^2=n_{i,\sigma}$~\cite{tremblay2011two}. The TPSC approach addresses this issue by employing distinct vertex functions for the spin and charge channels,
\begin{equation}
    \begin{split}
    \chi^{\mathrm{sp}} &= \frac{\chi^0}{1-\frac{1}{2}U_{\mathrm{sp}}\chi^0},\\
    \chi^{\mathrm{ch}} &= \frac{\chi^0}{1+\frac{1}{2}U_{\mathrm{ch}}\chi^0}, 
    \end{split}
    \label{Eq11}
\end{equation}
where $U_{\mathrm{sp}}$ and $U_{\mathrm{ch}}$ are in general different from the Hubbard interaction $U$. 
In this manner, additional degrees of freedom are introduced to ensure that the Pauli exclusion principle is satisfied. By substituting $n_{i,\sigma}^2=n_{i,\sigma}$ into the diagonal elements of Eq.~\eqref{Eq9}, we obtain
\begin{equation}
    \begin{split}
    \chi^{\mathrm{sp}}_{i,i}(t,t) &= 2\mathrm{i}\left<\hat{n}_{i,\uparrow}(t)\hat{n}_{i,\downarrow}(t)\right>-\mathrm{i}N_i(t),\\
    \chi^{\mathrm{ch}}_{i,i}(t,t) &= -2\mathrm{i}\left<\hat{n}_{i,\uparrow}(t)\hat{n}_{i,\downarrow}(t)\right>-\mathrm{i}N_i(t)+\mathrm{i}N_i^2(t),
    \end{split}
    \label{Eq12}
\end{equation}
where $N_i(t)=n_{i,\uparrow}(t)+n_{i,\downarrow}(t)$. These two equations are referred to as the sum rules for the spin and charge susceptibilities. However, an additional condition is required to close Eqs.~\eqref{Eq12} and \eqref{Eq11}, since the double occupation $\left<\hat{n}_{i,\uparrow}(t)\hat{n}_{i,\downarrow}(t)\right>$ remains unknown. In practice, it is common to adopt the following Ansatz, inspired by Ref.~\onlinecite{hedayati1989ground},
\begin{equation}
U\left<\hat{n}_{i,\uparrow}(t)\hat{n}_{i,\downarrow}(t)\right>=U_{\mathrm{sp}}\left<\hat{n}_{i,\uparrow}(t)\right>\left<\hat{n}_{i,\downarrow}(t)\right>.
\label{Eq13}
\end{equation}
Without considering this specific Ansatz, as long as the double occupation remains physical, which means $0 \leq \left<\hat{n}_{i,\uparrow}(t)\hat{n}_{i,\downarrow}(t)\right> \leq N_i^2/4$ for a repulsive $U$, the local spin and charge susceptibilities will always be finite. This ensures that the susceptibility does not diverge at any momentum in two-dimensional systems (note that this is not the case in three-dimensional systems~\cite{tremblay2011two}). In other words, the TPSC method satisfies the Mermin-Wagner theorem.

\begin{figure}[t]
    \includegraphics[width=\linewidth]{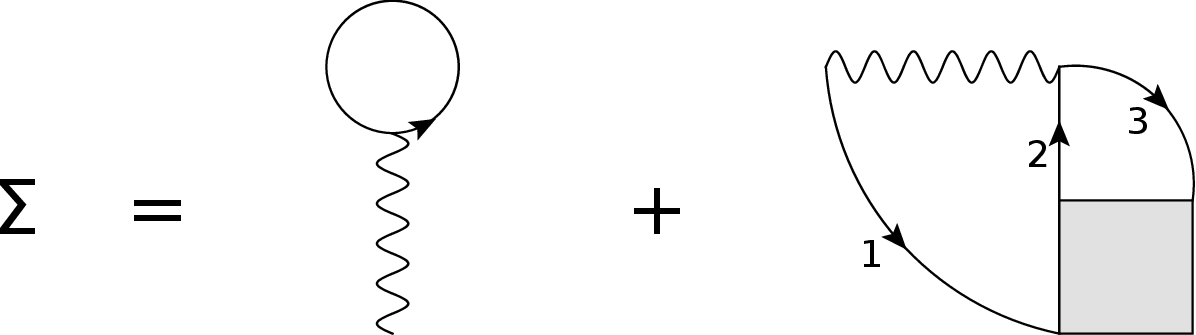}
    \caption{Feynman diagrams for the self-energy $\Sigma^{(2)}$ in Eq.~\eqref{Eq14}.  The wavy lines represent the bare interaction $U$, while the solid lines with arrows denote the electron propagators $G^{(1)}$. In the Hubbard model, the interaction $U$ acts only between electrons with opposite spins. Consequently, lines 2 and 3 must have the same spin, which is opposite to the spin of line 1. The gray box in the diagram represents the reducible vertex.}
    \label{fig1}
\end{figure}

After solving for the susceptibilities, the second step in TPSC is to obtain a better approximation for the self-energy. Since collective behavior is less influenced by single-particle properties, a ``one-shot" procedure is adopted: the susceptibilities are assumed to remain unaffected by the self-energy that is derived. To distinguish quantities from the first and second steps, we denote the former with a superscript $(1)$ and the latter with a superscript $(2)$. Using the Dyson equation, the exact self-energy can be expressed by Feynman diagrams, as shown in Fig.~\ref{fig1}. It can be divided into two parts: the Hartree term and the contribution from the fully reducible vertex $\Sigma^{C}$ (represented by the gray box in the figure): 
\begin{equation}
    \Sigma^{(2)}_{i,j}(t,t')=-\mathrm{i}UG^{(1)}_{i,i}(t,t)+\Sigma^{C}_{i,j}(t,t').
    \label{Eq14}
\end{equation}
To derive a more detailed expression, the reducible vertex must be expanded into irreducible vertices. This can be achieved through two approaches: the first involves expanding the component connected to two parallel-spin Green's functions (lines 2 and 3 in Fig.~\ref{fig1}), while the second involves expanding the component connected to two antiparallel-spin Green's functions (lines 1 and 2 in Fig.~\ref{fig1}) into ladder diagrams. These are generally referred to as the longitudinal channel and the transversal channel, respectively.

Since the contribution from the reducible vertex can be expressed as a functional derivative with respect to an external field, we can couple either a field along $z$ or $x$ to the Pauli matrices to derive the susceptibilities \cite{simard2023dynamical}. The self-energy $\Sigma^{C}_{i,j}(t,t')$ for the longitudinal channel can then be expressed as 
\begin{equation}
    \Sigma^{C,\mathrm{lo}}_{i,j}(t,t')=-\mathrm{i}\frac{U}{4}G^{(1)}_{i,j}(t,t')\left[U_{\mathrm{ch}}\chi^{\mathrm{ch}}_{j,i}(t',t)+U_{\mathrm{sp}}\chi^{\mathrm{sp}}_{j,i}(t',t)\right].
    \label{Eq15}
\end{equation}
Similarly, the self-energy $\Sigma^{C}_{i,j}(t,t')$ for the transversal channel becomes 
\begin{equation}
    \Sigma^{C,\mathrm{tr}}_{i,j}(t,t')=-\mathrm{i}\frac{U}{2}G^{(1)}_{i,j}(t,t')U_{\mathrm{sp}}\chi^{\mathrm{sp}}_{j,i}(t',t).
    \label{Eq16}
\end{equation}
In TPSC, we average the results from the two channels to obtain the following estimation of the self-energy:
\begin{equation}
    \Sigma^{C}_{i,j}(t,t')=-\mathrm{i}\frac{U}{8}G^{(1)}_{i,j}(t,t')\left[U_{\mathrm{ch}}\chi^{\mathrm{ch}}_{j,i}(t',t)+3U_{\mathrm{sp}}\chi^{\mathrm{sp}}_{j,i}(t',t)\right].
    \label{Eq17}
\end{equation}
It is important to stress that TPSC is not a perturbative diagrammatic method like FLEX, which we will introduce in Sec.~\ref{sec_flex}, so the factors in the self-energy expression are different in the two methods.

Once the self-energy $\Sigma^{(2)}$ has been obtained, we can calculate the corresponding Green's function $G^{(2)}$. In this step, the chemical potential is adjusted to impose a consistent density $N^{(2)}=N^{(1)}$. All the TPSC Green's functions shown in the results section are $G^{(2)}$.

Besides being invalid in the strong interaction regime, TPSC fails to perform well deep inside the renormalized classical regime~\cite{vilk1997non}, where the spin fluctuations become very strong. To address this issue, several variants of TPSC have been developed, including TPSC+, TPSC+SFM, and TPSC+GG~\cite{gauvin2023improved,simard2023dynamical}. The common idea behind these approaches is to iteratively feed the information from $G^{(2)}$ back into $G^{(1)}$ until self-consistency is achieved. Another approach to tackle this problem is to use the double occupation from alternative methods instead of the Ansatz \eqref{Eq13}. For instance, in TPSC+DMFT \cite{martin2023nonlocal,simard2023dynamical,zantout2023improved}, the DMFT framework provides a more accurate estimation of the double occupation, which is then utilized in TPSC. We will introduce the TPSC+DMFT method in the following subsection.

\subsection{TPSC+DMFT}

As mentioned in the previous subsection, the Ansatz \eqref{Eq13} becomes unreliable deep inside the renormalized classical regime. Furthermore, using this Ansatz can cause the charge susceptibility to diverge in the strong-interaction regime, leading to unphysical results. As an alternative, we can use the double occupation calculated within DMFT, which provides a more accurate estimation of local quantities. More specifically, when combining the TPSC method with DMFT, the first step is to substitute the double occupation obtained from DMFT into Eq.~\eqref{Eq12}, instead of relying on Eq.~\eqref{Eq13}, to solve for $U_{\mathrm{sp}}$ and $U_{\mathrm{ch}}$.

In the second step of TPSC, which involves the calculation of the self-energy, a similar substitution can also be applied. It is generally expected that DMFT provides a good estimate of the local self-energy. Therefore, we replace the local self-energy obtained within the TPSC framework by the DMFT result. In frequency and momentum space, the local and nonlocal self-energy components in TPSC can be expressed as
\begin{equation}
    \begin{split}
    \Sigma^{(2)}_{\mathrm{loc}}(\omega) &= \frac{1}{N_{\mathbf{k}}}\sum_{\mathbf{k}}\Sigma^{(2)}(\mathbf{k},\omega),\\
    \Sigma^{(2)}_{\mathrm{nonloc}}(\mathbf{k},\omega)&=\Sigma^{(2)}(\mathbf{k},\omega)-\Sigma^{(2)}_{\mathrm{loc}}(\omega).
    \end{split}
    \label{Eq18}
\end{equation}
From the perspective of DMFT, this self-energy substitution can also be interpreted as adding the above nonlocal self-energy into the DMFT loop. Since the DMFT self-energy is not explicitly calculated in the DMFT loop, we extract the nonlocal self-energy from TPSC and subtract it in the calculation of the inverse lattice Green's function (Eq.~\eqref{Eq6}). 

The incorporation of the TPSC nonlocal self-energy can slightly alter the original DMFT density and double occupation; however, these changes are negligible in practice. We therefore implement the TPSC+DMFT calculation in a one-shot fashion. First, a complete DMFT calculation is performed and then the DMFT-derived double occupation is used to calculate the spin and charge vertices within TPSC. Finally, the nonlocal self-energy from TPSC is incorporated into Eq.~\eqref{Eq6} to compute the final Green's function $G^{(2)}$, without further self-consistent iterations. We attempted to iterate the above procedure, as done in Ref.~\onlinecite{simard2023dynamical}, but the process failed to converge when using the NCA solver adopted in this study.

An alternative approach is to combine variants of TPSC, such as TPSC+GG, with DMFT. However, we observed that when the Green's function from DMFT is fed back into $G^{(1)}$ of TPSC, the converged results closely resemble the original DMFT results, and the nonlocal self-energy becomes negligible. We show some results in Appendix~\ref{app_A}. This behavior could also be linked to the limitations of the NCA solver.

Although TPSC+DMFT has been previously explored, these studies have primarily focused on the weak to moderate interaction regime~\cite{martin2023nonlocal,zantout2023improved}. In this work, we study the potential of TPSC+DMFT in the moderate to strong interaction regime, and for the description of doped Mott systems. The momentum- and frequency-resolved Green's functions offer useful insights into the strengths and limitations of this method.

\subsection{FLEX}
\label{sec_flex}

The FLEX approximation is commonly used for comparison with the TPSC approach, since both methods are rooted in the RPA framework. FLEX is based on the concept of conserving approximations proposed by Baym and Kadanoff~\cite{baym1962self,kadanoff2018quantum,bickers1989conserving}.

It satisfies the Mermin-Wagner theorem in the weakly correlated regime. However, 
%violates the Mermin-Wagner theorem~\cite{kitatani2015flex+} and, 
because the first step in FLEX -- the calculation of the susceptibilities -- is based on Eq.~\eqref{Eq10}, %the method violates 
also 
the Pauli principle \cite{vilk1997non}.

In the second step of FLEX, the self-energy is computed. As its name suggests, the method accounts for the exchange of collective fluctuations. Consequently, the self-energy is expressed as the sum of four distinct components:
\begin{equation}
    \Sigma^{\mathrm{FLEX}}=\Sigma^{\mathrm{HF}}+\Sigma^{\mathrm{SB}}+\Sigma^{\mathrm{ph}}+\Sigma^{\mathrm{pp}}.
    \label{Eq19}
\end{equation}
Here, $\Sigma^{\mathrm{HF}}$ represents the first-order term from the Hartree-Fock approximation, $\Sigma^{\mathrm{SB}}$ corresponds to the second-order term, $\Sigma^{\mathrm{ph}}$ accounts for the contribution from particle-hole fluctuations beyond second order, and $\Sigma^{\mathrm{pp}}$ represents the contribution from particle-particle fluctuations beyond second order. In this study, we consider only the first three contributions. The particle-hole fluctuation term, $\Sigma^{\mathrm{ph}}$, can be expressed as
\begin{equation}
    \begin{split}
\Sigma^{\mathrm{ph}}_{i,j}(t,t')=&-\mathrm{i}UG_{i,j}(t,t')\left\{\frac{U}{4}\left[\chi^{\mathrm{ch}}_{j,i}(t',t)-\chi^{\mathrm{0}}_{j,i}(t',t)\right]\right.\\
& \left.+\frac{3U}{4}\left[\chi^{\mathrm{sp}}_{j,i}(t',t)-\chi^{\mathrm{0}}_{j,i}(t',t)\right]\right\},
    \end{split}
    \label{Eq20}
\end{equation}
where $\chi^{\mathrm{ch}}$ and $\chi^{\mathrm{sp}}$ are defined as in Eq.~\eqref{Eq10}. However, unlike the one-shot TPSC approach, FLEX is a self-consistent method. Consequently, the Green's functions used to determine $\chi^0$ are not the bare Green's functions. Instead, $\chi^0_{i,j}(t,t')$ is given by $\chi^0_{i,j}(t,t')=-2\mathrm{i}G_{i,j}(t,t')G_{j,i}(t',t)$, where $G_{i,j}(t,t')$ is the Green's function incorporating self-energy corrections. In the FLEX framework, the Green's function is obtained from the self-energy, which is then iteratively substituted into $\chi^0$ until self-consistency is achieved.

\begin{figure}[t]
    \includegraphics[width=\linewidth]{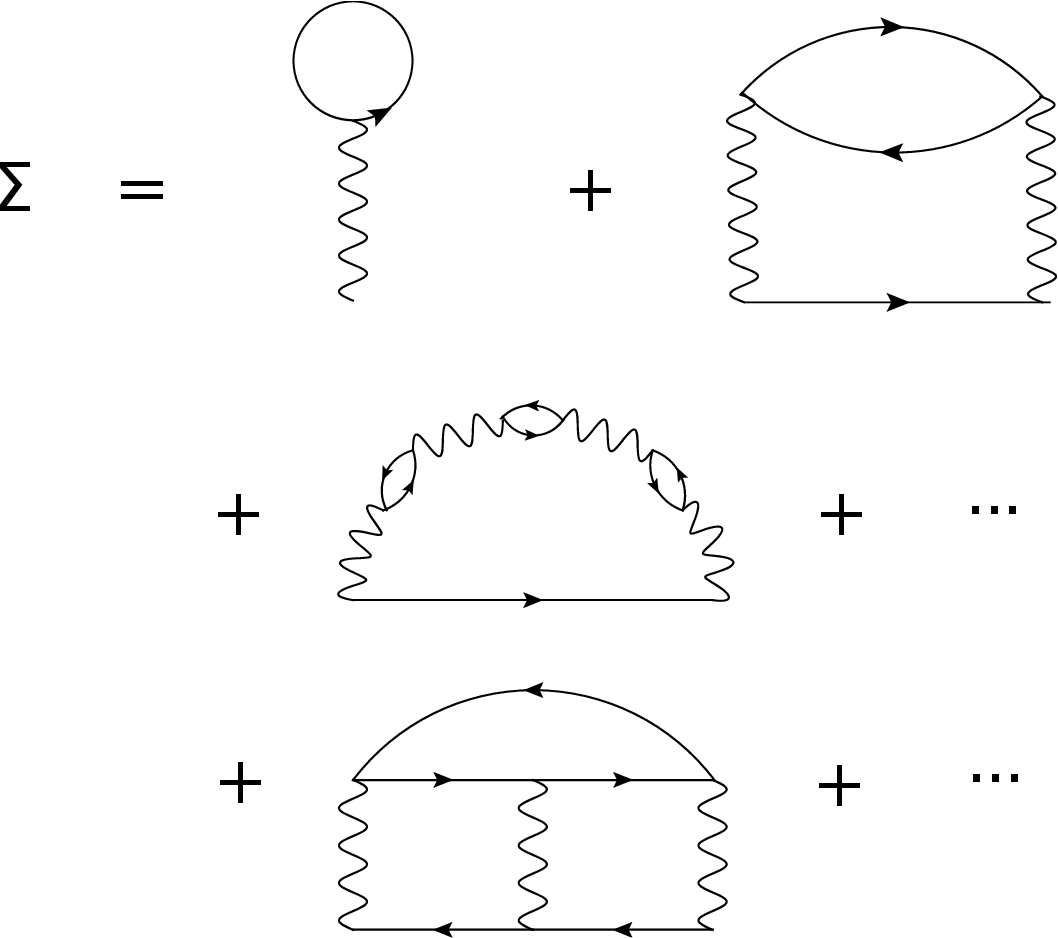}
    \caption{Feynman diagrams for the self-energy $\Sigma^{\mathrm{FLEX}}$. The wavy lines represent the bare interaction $U$, while the solid lines with arrows denote the full electron propagators $G$. The bubble diagrams in the second row include only those with an odd number of bubbles. The ladder diagrams in the third row, on the other hand, contain configurations with even and odd number of rungs.}
    \label{fig2}
\end{figure}

The Feynman diagrams for the self-energies in Eq.~\eqref{Eq20} are illustrated in Fig.~\ref{fig2}. The first row depicts the Hartree term and the second-order term. The bubble diagrams in the second row arise from the sum of the density fluctuation exchange ($S=0, S_z=0$) and longitudinal magnetic fluctuation exchange ($S=1, S_z=0$) contributions~\cite{bickers1989conserving}. Notably, the longitudinal magnetic fluctuations partially cancel the density fluctuations, leaving only bubble diagrams with an odd number of bubbles. The ladder diagrams in the third row originate from two transverse magnetic fluctuation channels ($S=1, S_z=\pm 1$). Since there are three magnetic channels and one density channel, the prefactor for $\chi^{\mathrm{sp}}$ is three times that for $\chi^{\mathrm{ch}}$.

It should be noted that FLEX differs from TPSC under the assumptions $U_{\mathrm{sp}} = U$ and $U_{\mathrm{ch}} = U$. Specifically, there is a difference in the prefactors of the higher-order terms, as TPSC involves an averaging and  FLEX a summation of diagram contributions.

FLEX is a perturbative approach suitable for the weak to moderate interaction regime. It has been applied in various contexts, such as spin-fluctuation-mediated superconductivity~\cite{arita2009fermi}. However, FLEX cannot accurately capture the physics of the pseudogap, unlike TPSC~\cite{moukouri2000many,vilk1997non}. Moreover, it is incapable of describing the Mott insulating state.

\subsection{Steady-state implementation}

In contrast to previous studies, our simulations are implemented directly on the real frequency axis. Consequently, the spectral function is calculated without the need for analytic continuation, eliminating a major source of uncertainties. Since we simulate a steady-state system, which is time-translation invariant, we only need to consider a two-branch Schwinger-Keldysh contour instead of the three-branch Kadanoff-Baym contour.

The general procedure for implementing the steady-state involves deriving the aforementioned equations for all relevant components in the real-time and real-space domains using Langreth's rules~\cite{stefanucci2013nonequilibrium,haug2008quantum}. These equations are then transformed into frequency and momentum space via fast Fourier transformation to facilitate the computations. Detailed explanations can be found in Refs.~\onlinecite{li2021,yan2024spin}. 

\begin{figure*}[t]
\includegraphics[width=\linewidth]{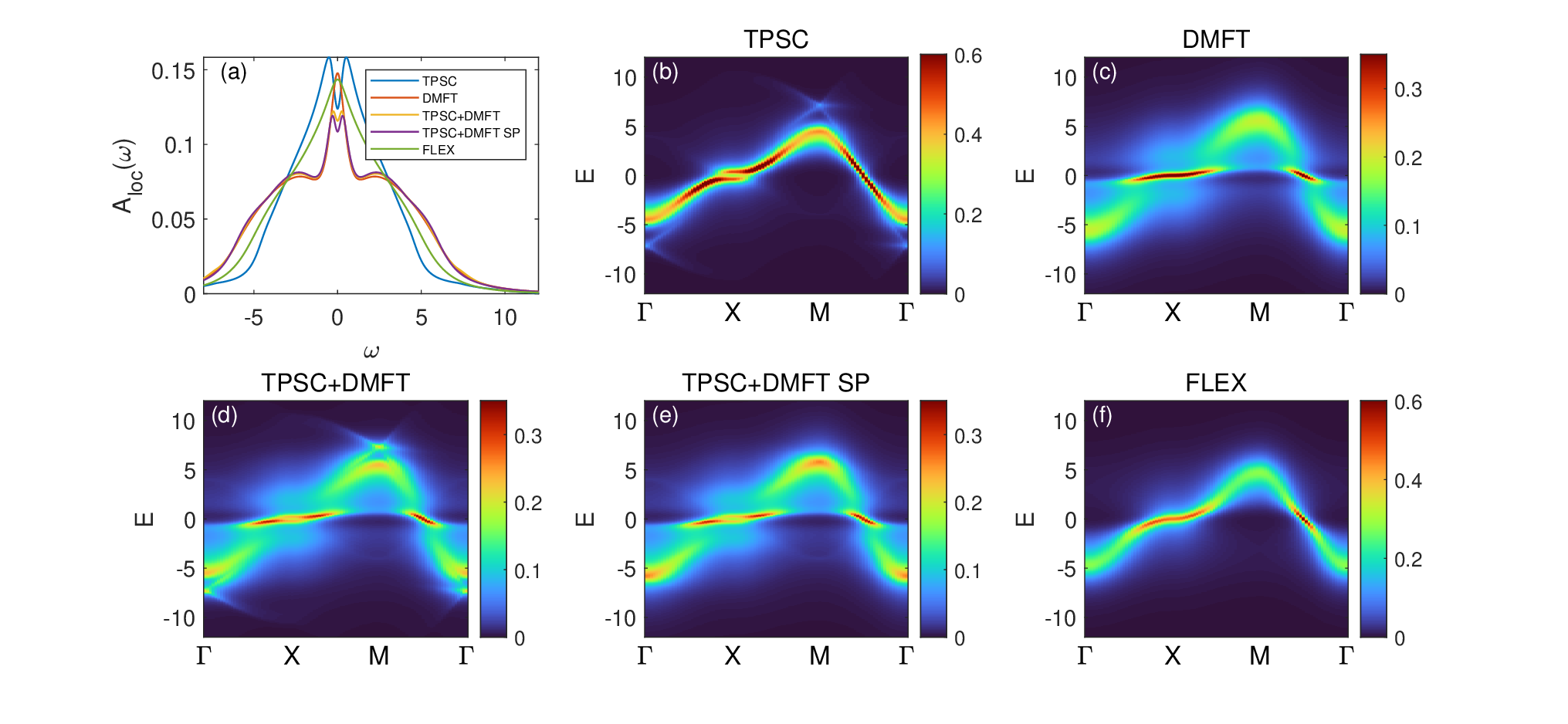}
\caption{Spectral functions for the half-filled Hubbard model with $U=4$ and $t'=0$ at temperature $T=0.25$. (a) Comparison of the local spectral functions obtained from the indicated methods. Panels (b)-(e) show the $\mathbf{k}$-resolved spectral functions from (b) TPSC, (c) DMFT, (d) TPSC+DMFT, and (e) TPSC+DMFT with the TPSC self-energy corresponding to the transversal channel in Eq.~\eqref{Eq16} only. (f) $\mathbf{k}$-resolved spectral function from FLEX.}
\label{fig3}
\end{figure*}

To stabilize the chemical potential and accelerate convergence, we incorporate a bath. In the TPSC calculations, the retarded component of the bath self-energy, $\Sigma^{\mathrm{B,r}}$, is given by
\begin{equation}
\Sigma^{\mathrm{B,r}}(\omega')=\begin{cases} 
    \frac{\Gamma}{D}\left(\omega'-\mathrm{i}\sqrt{D^2-\omega'^2}\right), & \left|\omega'\right|\le D, \\ 
    \frac{\Gamma}{D}\left[\omega'-\mathrm{sgn}(\omega')\sqrt{\omega'^2-D^2}\right], & \left|\omega'\right|> D.
    \end{cases}
    \label{Eq21}
\end{equation}
Here, $\omega' = \omega - \mu$, $D$ denotes the half bandwidth, and $\Gamma$ represents the coupling strength to the bath. The corresponding lesser component is given by $\Sigma^{\mathrm{B,<}}(\omega')=-2\mathrm{i}f(\omega')\mathrm{Im}\left[\Sigma^{\mathrm{B,r}}(\omega')\right]$, where $f(\omega')$ is the equilibrium Fermi-Dirac distribution. In this study, we employ a wide-bandwidth, weakly-coupled bath with parameters $D = 32$ and $\Gamma = 0.05$.

Uniform mesh parameters are used in all the calculations. The entire first Brillouin zone is divided into a $64 \times 64$ $\mathbf{k}$-grid, and the frequency interval is set to $d\omega = 0.001$, with a total of $2^{16}$ frequency points considered.

\section{Results}

\subsection{Moderately correlated half-filled system}\label{section3A}

First, we examine the half-filled model in the moderate interaction regime with $U = 4$. The temperature is set to $T = 0.25$, and only the nearest-neighbor hopping is considered. TPSC provides a good description of the two-dimensional Hubbard model in this regime \cite{vilk1997non} (except for the magnetic correlation length \cite{Schaefer2021,simard2023dynamical}). In particular, it has been demonstrated that, unlike FLEX, TPSC is capable of reproducing the pseudogap in the spectral function. The NCA solver employed in our DMFT calculations is however not ideally suited for describing moderately correlated metals, as it tends to underestimate the double occupation~\cite{eckstein2010nonequilibrium}. This could be compensated by reducing $U$ in the DMFT calculations. To avoid adjustable parameters, we however prefer to use the same $U$ as in TPSC, keeping in mind the bias from the NCA. Recently developed higher-order steady-state solvers~\cite{Eckstein2024,Kim2024} will eliminate this uncertainty. 

\subsubsection{Spectral functions}

The local and $\mathbf{k}$-resolved spectral functions obtained with the various discussed methods are presented in Fig.~\ref{fig3}. The path for the $\mathbf{k}$-resolved spectra connects the high-symmetry points in the first Brillouin zone: $\Gamma = (0,0) \rightarrow \text{X} = (\pi,0) \rightarrow \text{M} = (\pi,\pi) \rightarrow \Gamma = (0,0)$.

All the TPSC results shown in this figure and the figures below refer to TPSC combined with the double occupation from DMFT, but without the local self-energy substitution. This is because TPSC combined with the original Ansatz in Eq.~\eqref{Eq13} lacks solutions for $U_{\mathrm{sp}}$ or $U_{\mathrm{ch}}$ for most parameter sets considered in this paper. Also, as mentioned before, DMFT is expected to provide a better estimate of the double occupation. In the case shown in Fig.~\ref{fig3}(b), the substitution of the double occupation only affects $U_{\mathrm{ch}}$, changing its value from 13.31 to 17.80, while $U_{\mathrm{sp}}=2.17$ is unchanged. In this panel, we observe a clear suppression of spectral weight near the X point (antinode), leading to the pseudogap in the local spectral function shown in Fig.~\ref{fig3}(a), while a metallic band crosses the Fermi level half-way between M and $\Gamma$ (nodal point $\mathbf{k}=(\frac{\pi}{2},\frac{\pi}{2})$). Additionally, high-energy structures appear near the M and $\Gamma$ points. These features are related to $\chi^{\mathrm{ch}}$ and will be discussed below. Note that neither the pseudogap nor the charge susceptibility related features are present in the FLEX result shown in Fig.~\ref{fig3}(f).

\begin{figure*}[t]
\includegraphics[width=\linewidth]{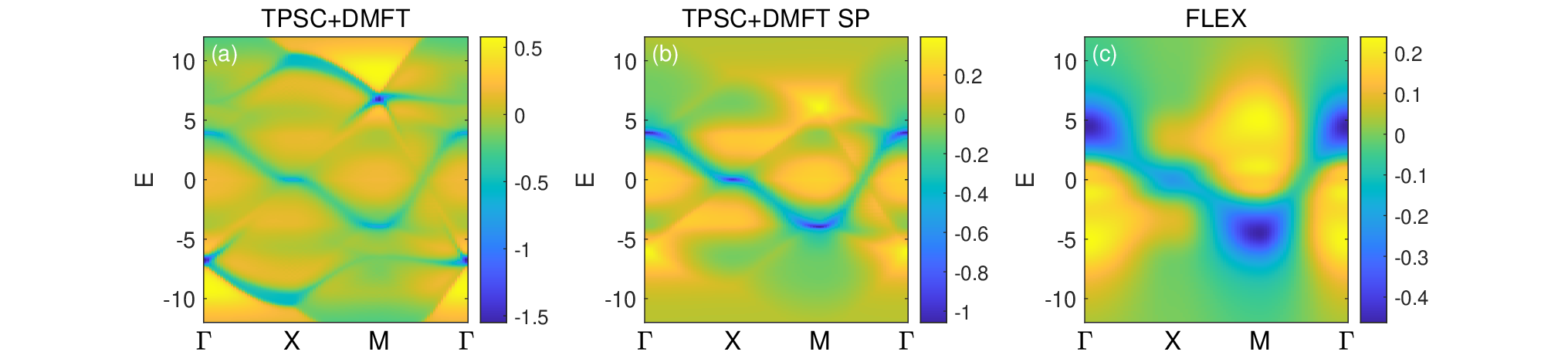}
\caption{Imaginary part of the nonlocal self-energy as a function of momentum and energy for the half-filled Hubbard model with $U=4$ and $t'=0$ at temperature $T=0.25$. Panels (a)-(c) show the results for (a) TPSC+DMFT (same as TPSC), (b) TPSC+DMFT with the TPSC self-energy corresponding to the transversal channel in Eq.~\eqref{Eq16} only, and (c) FLEX.}
\label{fig4}
\end{figure*}

Figure~\ref{fig3}(c) presents the momentum-resolved DMFT spectral function. Comparing with the results from Ref.~\onlinecite{chatzieleftheriou2024local}, we find that the NCA solver overestimates the correlation strength, as expected, but the spectra are qualitatively consistent. They show a narrow quasi-particle band between two incoherent Hubbard bands which are separated by the energy $U$. There is no gap opening at X and no evidence for a pseudogap in the local spectral function (panel (a)), consistent with the general understanding that the pseudogap originates from nonlocal antiferromagnetic correlations \cite{Gunnarsson2015}. Since the interaction $U$ is only half the bandwidth, there is no Mott gap in this system. 

It is interesting to note that the DMFT spectral function looks like the superposition of a weakly and strongly renormalized band, with the former featuring prominent weight near $\Gamma$ and M, and the latter (which contributes the Hubbard satellites) dominating the weight near the X point. This is consistent with the effective multi-orbital description of the single-orbital Hubbard model \cite{werner2024}, which classifies the electrons into weakly and strongly correlated types, with the density of states of the strongly correlated ones concentrated in the antinodal region. 

When we add the nonlocal self-energy from TPSC to the local DMFT self-energy and perform a one-shot calculation, we get the spectral function shown in Fig.~\ref{fig3}(d). Now the spectral weight of the narrow quasi-particle band is suppressed near the X point, resulting in the small pseudogap evident in Fig.~\ref{fig3}(a). We also notice the TPSC-like high-energy structures near the $\Gamma$ and M points. If only the nonlocal TPSC self-energy from the transversal channel is added, the latter structures disappear, as shown in Fig.~\ref{fig3}(e), which indicates that they originate from charge fluctuations. On the other hand, the pseudogap persists, indicating that it is a consequence of nonlocal spin fluctuations.

The nonlocal self-energy from TPSC not only opens the pseudogap in the local spectral function, as shown in Fig.~\ref{fig3}(a), but also produces a faint mirror-like copy of the (non-interacting) electronic dispersion, symmetric about zero energy, as observed in Figs.~\ref{fig3}(d) and~\ref{fig3}(e). This feature reflects the system's proximity to the antiferromagnetic state with a doubled unit cell and has also been reported in D-TRILEX calculations~\cite{chatzieleftheriou2024local}. 

Overall, the TPSC+DMFT spectra show the features expected from nonlocal correlations, and clearly represent an improvement over the DMFT description. At the same time, they exhibit the strong quasi-particle renormalization and Hubbard band features which are characteristic of strongly correlated metals, and which cannot be captured by plain TPSC or FLEX.

\subsubsection{Self-energies and susceptibilities}

To explore how the nonlocal self-energy improves the results of DMFT, we plot the nonlocal self-energy obtained from TPSC and FLEX in Fig.~\ref{fig4}. Note that TPSC and TPSC+DMFT share the same nonlocal self-energy, while the DMFT self-energy is momentum independent. 

The nonlocal self-energy shown in Fig.~\ref{fig4}(a) corresponds to the average of the transversal and longitudinal channels, while Fig.~\ref{fig4}(b) presents the transverse contributions alone. By comparing these two figures, we identify poles with large negative values resulting from charge fluctuations: near $E = 7$ at the M point and $E = -7$ at the $\Gamma$ point. These poles are responsible for the high-energy structures in the single-particle spectral functions discussed earlier. Additionally, both figures reveal a clear band with a large negative nonlocal self-energy, mirrored with respect to the non-interacting band dispersion \eqref{Eq2}. This feature is the origin of the pseudogap and the faint mirror-like copy observed in the previous subsection. Studies based on exact diagonalization~\cite{eder2011self} and D-TRILEX~\cite{chatzieleftheriou2024local} also reported such a mirrored nonlocal self-energy, lending credibility to our findings.

The nonlocal self-energy from FLEX, shown in Fig.~\ref{fig4}(c), is qualitatively consistent with the TPSC result in reproducing the mirrored band. However, the pole near the X point in FLEX is not strong enough to form the pseudogap. More generally, the FLEX self-energy, which results from self-consistent iterations, is smoother than the one-shot TPSC self-energy contributions. 

\begin{figure}[t]
\includegraphics[width=\linewidth]{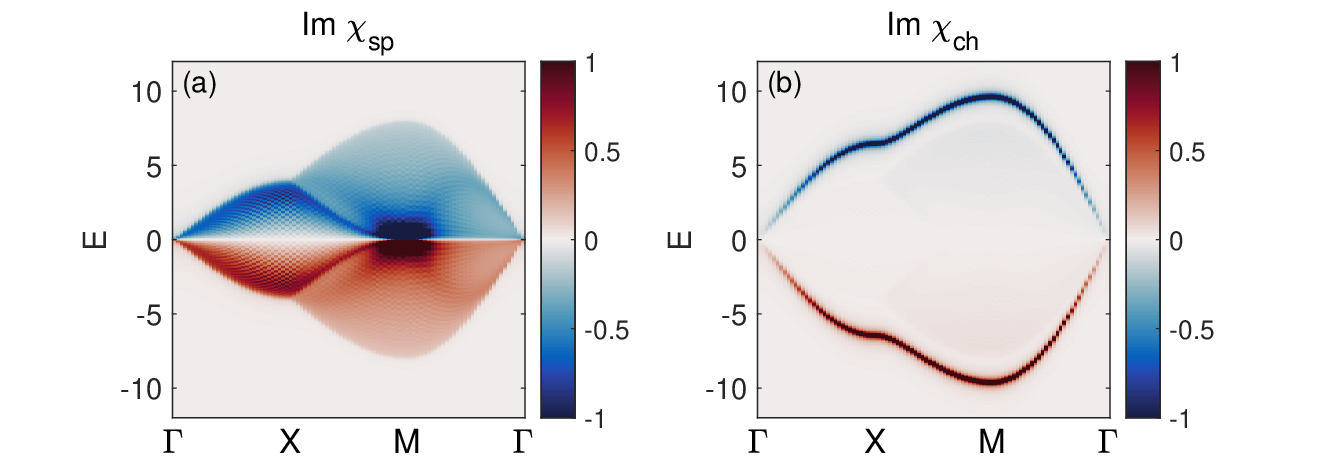}
\caption{Imaginary parts of the susceptibilities as a function of momentum and energy, for the half-filled Hubbard model with $U=4$ and $t'=0$ at temperature $T=0.25$. (a) $\mathrm{Im}\chi^{\mathrm{sp}}$ (spin susceptibility) and (b) $\mathrm{Im}\chi^{\mathrm{ch}}$ (charge susceptibility).
}
\label{fig5}
\end{figure}

Since the TPSC self-energy in Fig.~\ref{fig4} comes from the convolution of the non-interacting Green's function $G^{(1)}$ and $\chi^{\mathrm{sp/ch}}$, we plot the imaginary parts of the susceptibilities in Fig.~\ref{fig5} to gain further insights into the features of the nonlocal self-energy.

From Fig.~\ref{fig5}(a), we observe that the highest absolute values of $\mathrm{Im}\chi^{\mathrm{sp}}$ are reached near $E=0$ at the M point. Since the momenta and frequencies are summed during the convolution, we can attribute the features of the self-energy to various influences from $G^{(1)}$ and $\chi^{\mathrm{sp/ch}}$. For instance, $\Sigma^{C,\mathrm{tr}}$ along the momentum path from $\text{M}=(\pi,\pi)$ to $(2\pi,\pi)$ (or equivalently the path from $\text{M}=(\pi,\pi)$ to $\text{X}=(\pi,0)$) is determined by the contribution from $G^{(1)}$ along the path from $\Gamma=(0,0)$ to $\text{X}=(\pi,0)$ and $\chi^{\mathrm{sp}}$ at the M point. As the peak of $\chi^{\mathrm{sp}}$ at M is near $E=0$, the band of $\Sigma^{C,\mathrm{tr}}$ from $\text{M}=(\pi,\pi)$ to $\text{X}=(\pi,0)$ looks like the non-interacting energy band along $\Gamma=(0,0)$ to $\text{X}=(\pi,0)$. Given the particle-hole symmetry of the half-filled system with $t'=0$, a flipped energy band appears in the self-energy.

The high-energy structures in the spectral function can be similarly analyzed. The satellite at the $\text{M}=(\pi,\pi)$ point arises from the convolution of the flat band of $\chi^{\mathrm{ch}}$ near $E=7$, $\mathbf{k}=(0,\pi)$, and the flat band of $G^{(1)}$ near $E=0$, $\mathbf{k}=(\pi,0)$, or vice versa in momentum space. This can be interpreted as a process where an electron is inserted with momentum $(\pi,0)$ and simultaneously emits a plasmon with momentum $(0,\pi)$. Since this satellite is related to charge fluctuations, it is absent in Fig.~\ref{fig3}(e), which only considers the spin contributions. 

\subsubsection{Fermi surfaces}

\begin{figure}[t]
\includegraphics[width=\linewidth]{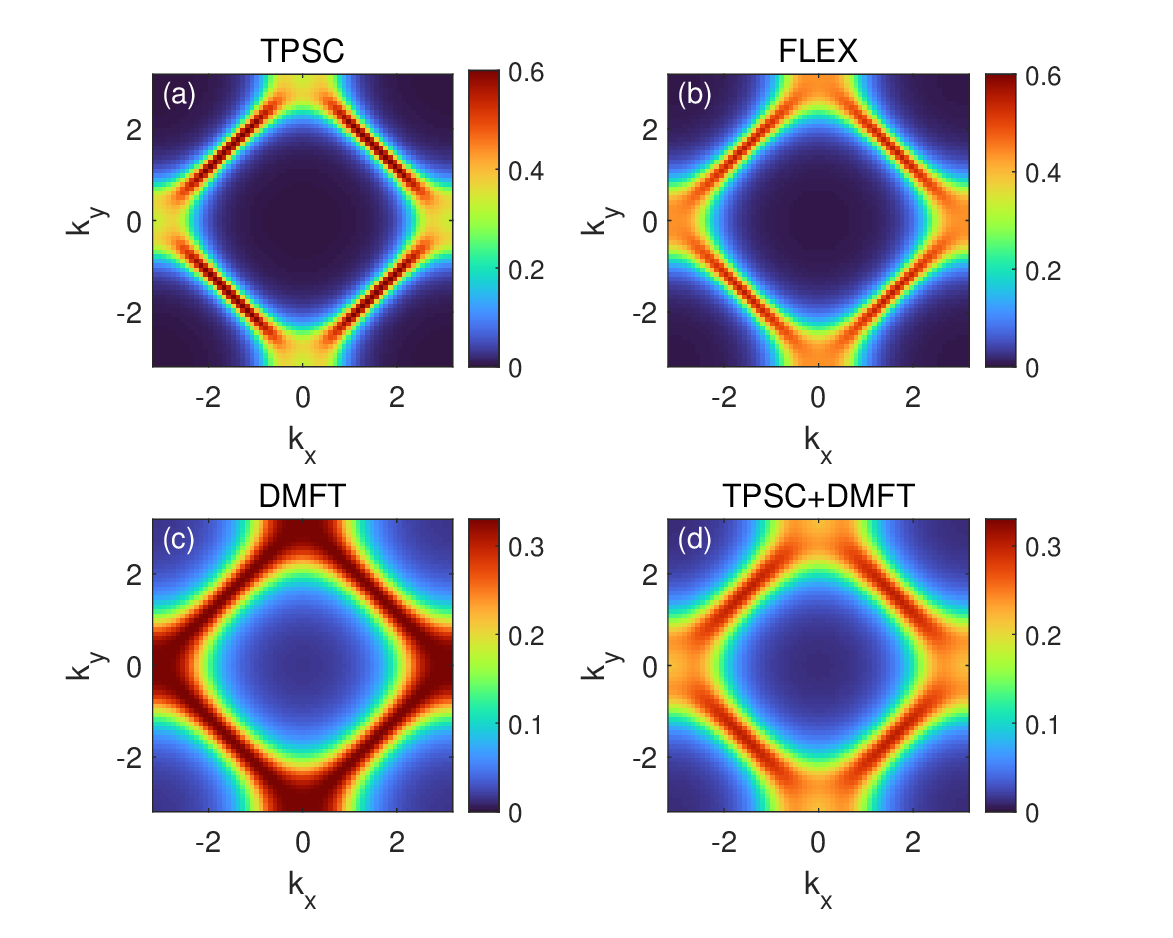}
\caption{Fermi surface of the half-filled Hubbard model with $U=4$ and $t'=0$ at temperature $T=0.25$, represented by the spectral function $A(\mathbf{k},\omega)$ averaged over $\omega\in [-0.2,0.2]$. Panels (a)-(d) depict the results within the first Brillouin zone obtained using (a) TPSC, (b) FLEX, (c) DMFT, and (d) TPSC+DMFT.}
\label{fig6}
\end{figure}

Next, we focus on the electronic structure near the Fermi level. The $\mathbf{k}$-resolved spectral functions averaged over $\omega \in [-0.2, 0.2]$ are presented in Fig.~\ref{fig6}. In the non-interacting limit, the Fermi surface forms a square connecting the four X points in the first Brillouin zone.

In the TPSC result shown in Fig.~\ref{fig6}(a), the suppression of spectral weight near the antinodal points leads to a Fermi surface composed of disconnected segments. This is related to the pseudogap opening by spin fluctuations discussed in connection with Fig.~\ref{fig3}. The nodal points ($\mathbf{k}=(\pm \frac{\pi}{2},\pm \frac{\pi}{2})$) are relatively unaffected by spin fluctuations. The FLEX results, shown in Fig.~\ref{fig6}(b), exhibit a similar pattern, but the suppression of weight at the antinodal points is significantly weaker.  

In contrast, the DMFT result in Fig.~\ref{fig6}(c) displays a continuous Fermi surface without momentum dependence of the quasi-particle lifetime. This is a consequence of the momentum-independent self-energy, or absence of spatial fluctuations, in the DMFT framework. Moreover, the Fermi surface in DMFT appears broader, and the peak intensity is lower compared to the TPSC and FLEX results, indicating stronger correlation effects. The TPSC+DMFT result in Fig.~\ref{fig6}(d) combines the features from both methods. On the one hand, the Fermi surface width and peak intensity are similar to those in the DMFT results. On the other hand, there is a clear suppression of spectral weight near the antinodal points, as in the TPSC results. The TPSC+DMFT Fermi surface is qualitatively consistent with results deduced from computationally much more heavy approaches like cluster DMFT or D$\Gamma$A \cite{Werner2009,rohringer2018}. This, together with the high momentum resolution and absence of analytical continuation, highlights the strengths of our real-frequency TPSC+DMFT approach.

\begin{figure*}[t]
\includegraphics[width=\linewidth]{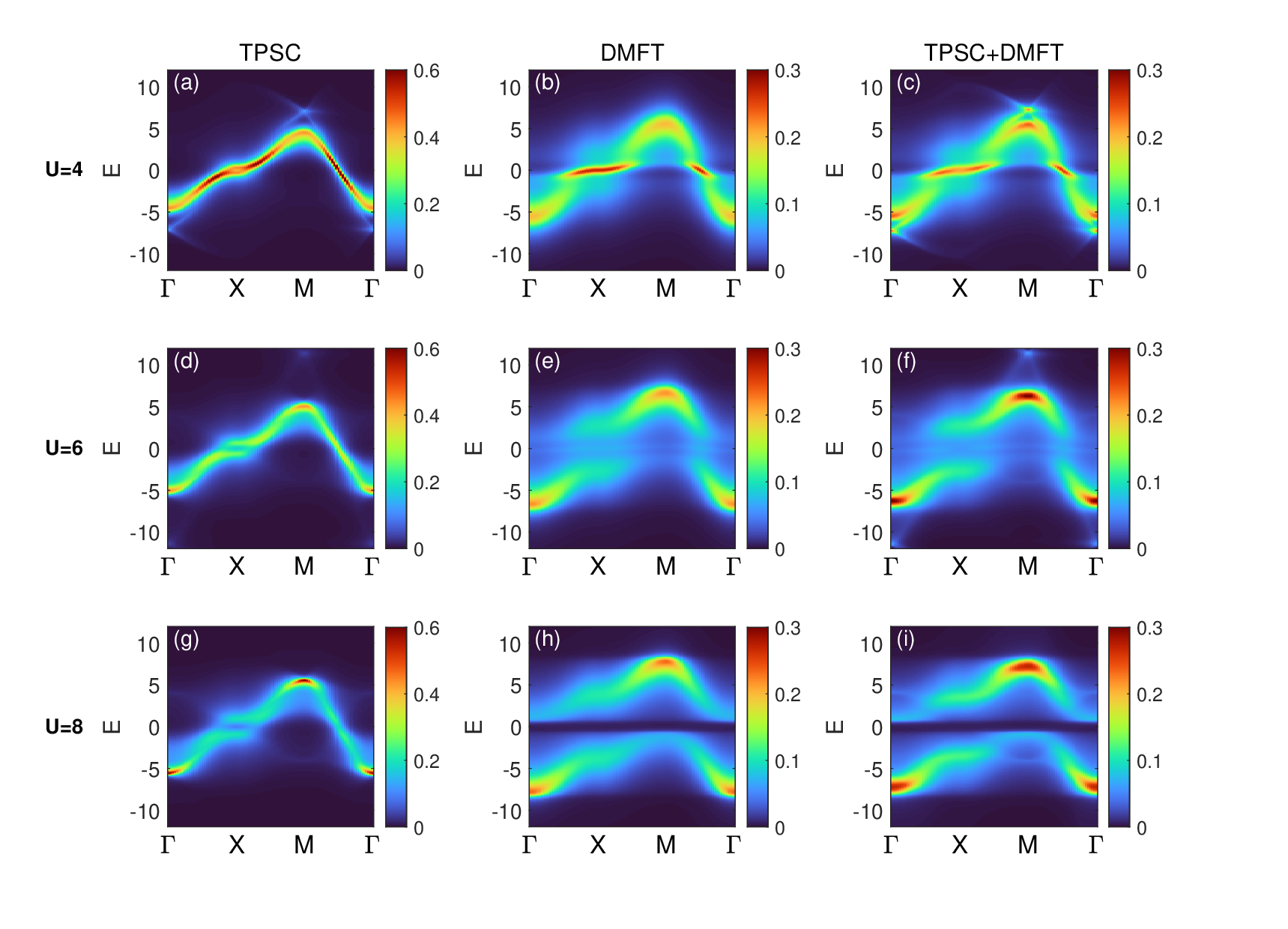}
\caption{Spectral functions for the half-filled Hubbard model with $t'=0$ at temperature $T=0.4$. Results obtained using the same method are grouped into columns, while the rows correspond to different interaction strengths $U$. Panels (a)-(c) are for $U=4$, (d)-(f) for $U=6$, and (g)-(i) for $U=8$. The first row (panels (a), (d), (g)) shows results obtained from TPSC, the second row (panels (b), (e), (h)) from DMFT, and the third row (panels (c), (f), (i)) from TPSC+DMFT.}
\label{fig7}
\end{figure*}

\subsection{Strongly correlated half-filled system}\label{section3B}

We now turn to half-filled systems with stronger interactions $U$, where TPSC is generally considered to be unreliable. Previous studies have reported significant deviations for $U>4$ in vertices and susceptibilities, when compared with more accurate methods~\cite{martin2023nonlocal,zantout2023improved}. Additionally, the vertices can sometimes exhibit unphysical behavior. While incorporating the double occupation from DMFT into TPSC can produce physical results, their validity remains questionable.

Initially, we performed calculations for $U=6$ and $U=8$ at $T=0.25$. However, the spin correlation length in TPSC is close to or beyond a diverging point at this low temperature, consistent with the so-called normalized classical regime~\cite{vilk1997non}. To mitigate the errors associated with the TPSC solution in this regime, we increase the temperature to $T=0.4$ in the subsequent calculations. Another point worth noting is that for $U=6$ and $U=8$, there may exist two pairs of solutions for $U_{\mathrm{sp}}$ and $U_{\mathrm{ch}}$. We select the more physical solution that satisfies $U_{\mathrm{sp}}<U$ and continuously connects to the solution for $U=4$. The resulting $\mathbf{k}$-resolved spectral functions from TPSC, DMFT, and TPSC+DMFT for $U=4$, $U=6$, and $U=8$ are shown in Fig.~\ref{fig7}.

To examine the temperature effects, we can compare the $U=4$, $T=0.4$ spectra in Figs.~\ref{fig7}(a)-\ref{fig7}(c) to the corresponding $T=0.25$ data Figs.~\ref{fig3}(b)-\ref{fig3}(d). In the TPSC case (Fig.~\ref{fig7}(a)), the pseudogap weakens significantly as the temperature increases, while the peak intensity of the DMFT result (Fig.~\ref{fig7}(b)) decreases. In the TPSC+DMFT spectrum (Fig.~\ref{fig7}(c)), the pseudogap becomes nearly indiscernible, indicating the pronounced thermal sensitivity of this feature.

\begin{figure*}[t]
\includegraphics[width=\linewidth]{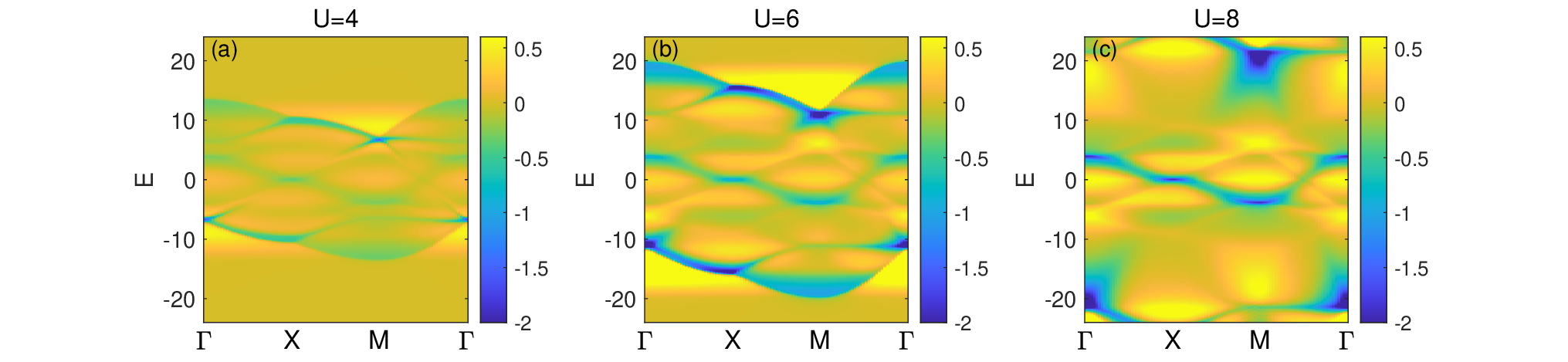}
\caption{Imaginary parts of the nonlocal self-energies for the half-filled Hubbard model with $t'=0$ at temperature $T=0.4$. Panel (a) shows the result for $U=4$, (b) for $U=6$, and (c) for $U=8$.}
\label{fig8}
\end{figure*}

\begin{table}
\centering
\begin{tabular}{|c|c|c|c|}
\hline
\diagbox{vertex}{interaction} & $U=4$ & $U=6$ & $U=8$ \\
\hline
$U^{\mathrm{sp}}$ & 2.21 & 2.49 & 2.61 \\
\hline
$U^{\mathrm{ch}}$ & 17.73 & 69.09 & 269.89 \\
\hline
\end{tabular}
\caption{The values of the vertices $U^{\mathrm{sp}}$ and $U^{\mathrm{ch}}$ calculated by TPSC (with DMFT double occupation) for different interactions at half filling.}
\label{table1}
\end{table}

In the TPSC spectra, the pseudogap becomes more pronounced as $U$ increases. As mentioned before, the formation of the pseudogap is closely related to the spin fluctuations, which can be characterized by the spin susceptibility. As the interaction strength increases, the vertices $U^{\mathrm{sp}}$ and $U^{\mathrm{ch}}$ change as reported in Table~\ref{table1}, and the system exhibits stronger antiferromagnetic correlations~\cite{Schaefer2021}. Measurements of the static spin susceptibility $\chi^{\mathrm{sp}}(\omega=0,\mathbf{k})$ at $\mathbf{k}=(\pi,\pi)$ yield the values $-2.93$, $-4.87$, and $-6.94$ for $U=4$, $U=6$, and $U=8$, respectively. Besides the enhancement of the pseudogap, the TPSC results also show a shift of spectral weight towards the edges of the spectrum, resembling the behavior seen in the DMFT spectra shown in Figs.~\ref{fig7}(b), \ref{fig7}(e) and \ref{fig7}(h). The feature at the M point associated with charge fluctuations is pushed to higher energies with increasing $U$, consistent with the evolution of the charge susceptibility dispersion. 

As expected, TPSC fails to describe the Mott transition; the system remains metallic even for 
$U=8$. In contrast, the DMFT result in Fig.~\ref{fig7}(h) represents a Mott insulating state. With increasing $U$, the narrow quasi-particle band in the DMFT results shrinks until a Mott gap forms, resulting in two distinct bands with similar dispersion. The lower band, being fully occupied, and the upper band, nearly empty, indicate that (up to quantum fluctuations) each site is singly occupied, representing a condition close to the atomic limit. The NCA solver employed here is expected to provide qualitatively correct results in this regime.

The TPSC+DMFT spectra, shown in Figs.~\ref{fig7}(d), \ref{fig7}(f), and \ref{fig7}(i), also display the Mott transition at large $U$. These results exhibit a behavior quite similar to the DMFT solutions, although the TPSC derived satellite at the M point persist and the weak quasi-particle band in the $U=6$ solutions shows a gap opening near the X point. In the Mott regime, the nonlocal self-energy contributions from TPSC do not significantly affect the robust Mott gap, but only slightly redistribute the spectral weight. In fact, the pseudogap physics becomes irrelevant in this regime, and the antiferromagnetic correlations merely lead to a slightly enhanced gap near the X point. 

To further elucidate the results shown in Fig.~\ref{fig7}, we present the nonlocal self-energies for the three TPSC+DMFT calculations in Fig.~\ref{fig8}. The data reveal that the intensity of the mirrored band of the self-energy near the Fermi surface increases as $U$ increases. This trend explains the enhanced pseudogap observed for larger $U$. Additionally, the contribution from $\chi^{\mathrm{ch}}$, which is responsible for the satellite structures in the spectral functions, shifts farther away from the contributions of $\chi^{\mathrm{sp}}$ near the Fermi surface as $U$ increases. While the intensity of this component does not diminish, the lack of energy overlap prevents it from influencing the spectral function. This behavior aligns with two-dimensional $t$-$J$ model calculations~\cite{putikka1994indications,tohyama1996approximate}, which provide an effective description of the half-filled Hubbard model in the large-$U$ limit.

Note that the appearance of the Hubbard bands manifests itself in the energy dependence of the local self-energy component, which is subtracted in Fig.~\ref{fig8}.  

\subsection{Doped Mott insulator ($t'=-0.3t$)}\label{section3C}

The single-band Hubbard model on the square lattice has been studied extensively as a minimal model for cuprate high-$T_c$ superconductors~\cite{monien1991application,avella2001two,feiner1996effective}. Superconductivity in these materials emerges out of a non-Fermi liquid metal state, upon doping the Mott insulating parent compounds. In this section, we investigate how well TPSC+DMFT describes the square lattice Hubbard model in the parameter regime relevant for cuprate physics~\cite{ferrero2009pseudogap,civelli2005dynamical,kyung2006pseudogap}. We switch on a next-nearest neighbor hopping $t'=-0.3$, fix the interaction to $U=8$ and study the effect of hole doping. The temperature is set to $T=0.05$.

The TPSC approach has been successfully used to study the effects of spin fluctuations on the Fermi surface in electron-doped cuprates with interactions in the range of $U=5\sim6$~\cite{kyung2004pseudogap,gauvin2022resilient}. In particular, this method has reproduced and explained key features observed in angle-resolved photoemission spectroscopy (ARPES) experiments. However, as discussed earlier, TPSC alone cannot describe the Mott insulating state at half-filling. We aim to overcome these limitations by applying the TPSC+DMFT method to doped Mott systems. While the accuracy of the TPSC nonlocal self-energy may be questionable in the case of the half-filled Mott insulator, it is expected to perform better in doped systems with weaker correlations.

\begin{figure*}[t]
\includegraphics[width=\linewidth]{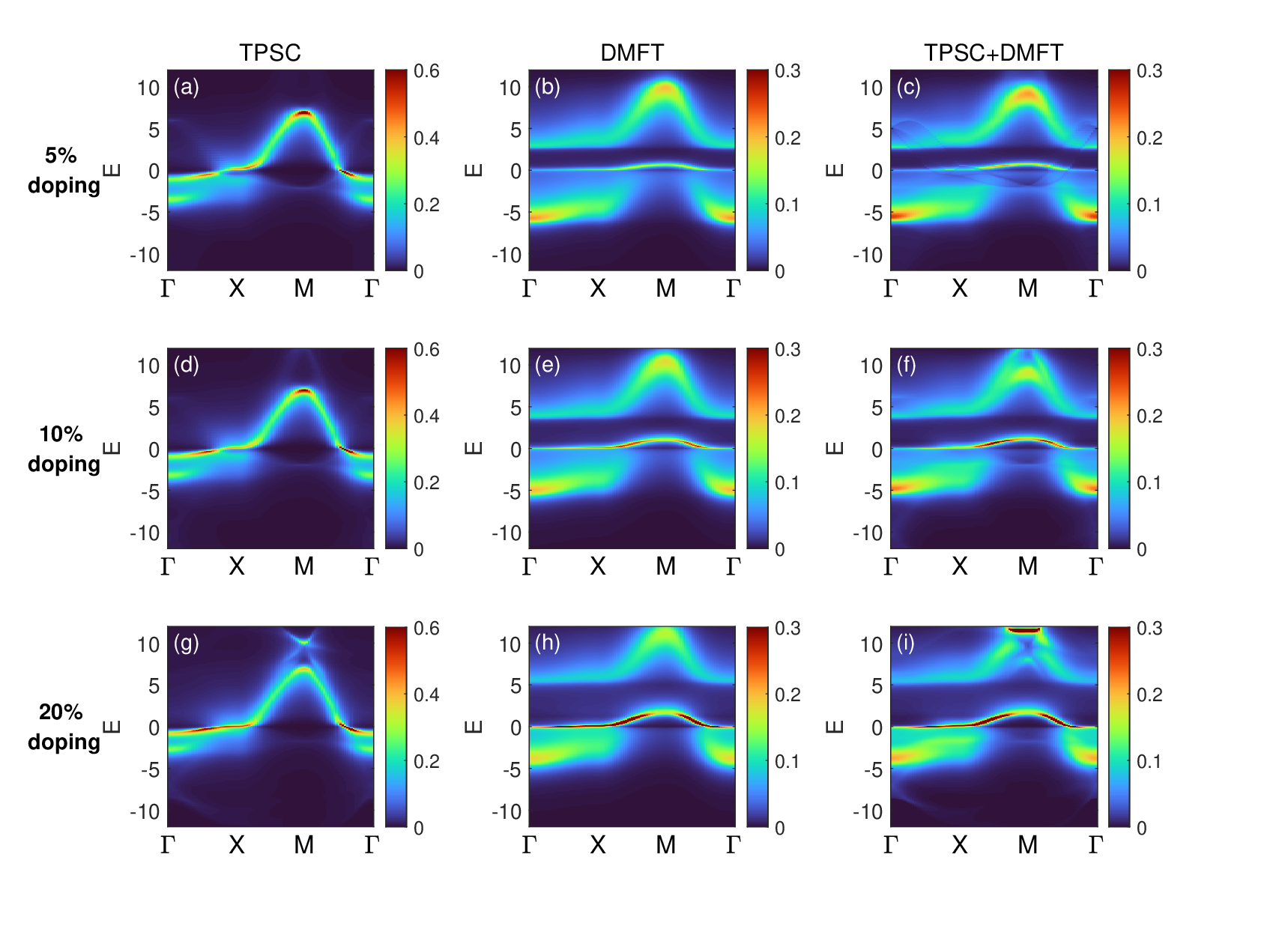}
\caption{Spectral functions for the hole-doped Hubbard model with $t'=-0.3t$ at temperature $T=0.05$. Results obtained using the same method are grouped into columns, while the rows correspond to different doping levels. Panels (a)-(c) represent $5\%$ doping, (d)-(f) $10\%$ doping, and (g)-(i) $20\%$ doping. The results in the first row (panels (a), (d), (g)) are from TPSC, those in the second row (panels (b), (e), (h)) from DMFT, and those in the third row (panels (c), (f), (i)) from TPSC+DMFT.
}
\label{fig9}
\end{figure*}

\subsubsection{Spectral functions}

\begin{table}
\centering
\begin{tabular}{|c|c|c|c|}
\hline
\diagbox{vertex}{doping level} & $5\%$ & $10\%$ & $20\%$ \\
\hline
$U^{\mathrm{sp}}$ & 2.51 & 2.43 & 2.26 \\
\hline
$U^{\mathrm{ch}}$ & 134.88 & 75.80 & 32.53 \\
\hline
\end{tabular}
\caption{The values of vertices $U^{\mathrm{sp}}$ and $U^{\mathrm{ch}}$ calculated by TPSC (with DMFT double occupation) for different hole doping levels at $U=8$.}
\label{table2}
\end{table}

We consider different hole-doping levels of $5\%$, $10\%$, and $20\%$. The spin and charge vertices calculated by TPSC decrease with increasing hole doping, as shown in Table~\ref{table2}. The spectral functions along the indicated path connecting high-symmetry points, calculated using TPSC, DMFT, and TPSC+DMFT, are shown in Fig.~\ref{fig9}. Comparing the results within the same column reveals that the satellite structure near the M point, which is very faint in the half-filled case, reappears at higher doping levels in both the TPSC and TPSC+DMFT spectra. Additionally, a narrow quasi-particle band emerges and gains weight with increasing doping in the DMFT and TPSC+DMFT results, indicating the appearance of a metallic state with strongly renormalized quasi-particles. 

\begin{figure*}[t]
\includegraphics[width=\linewidth]{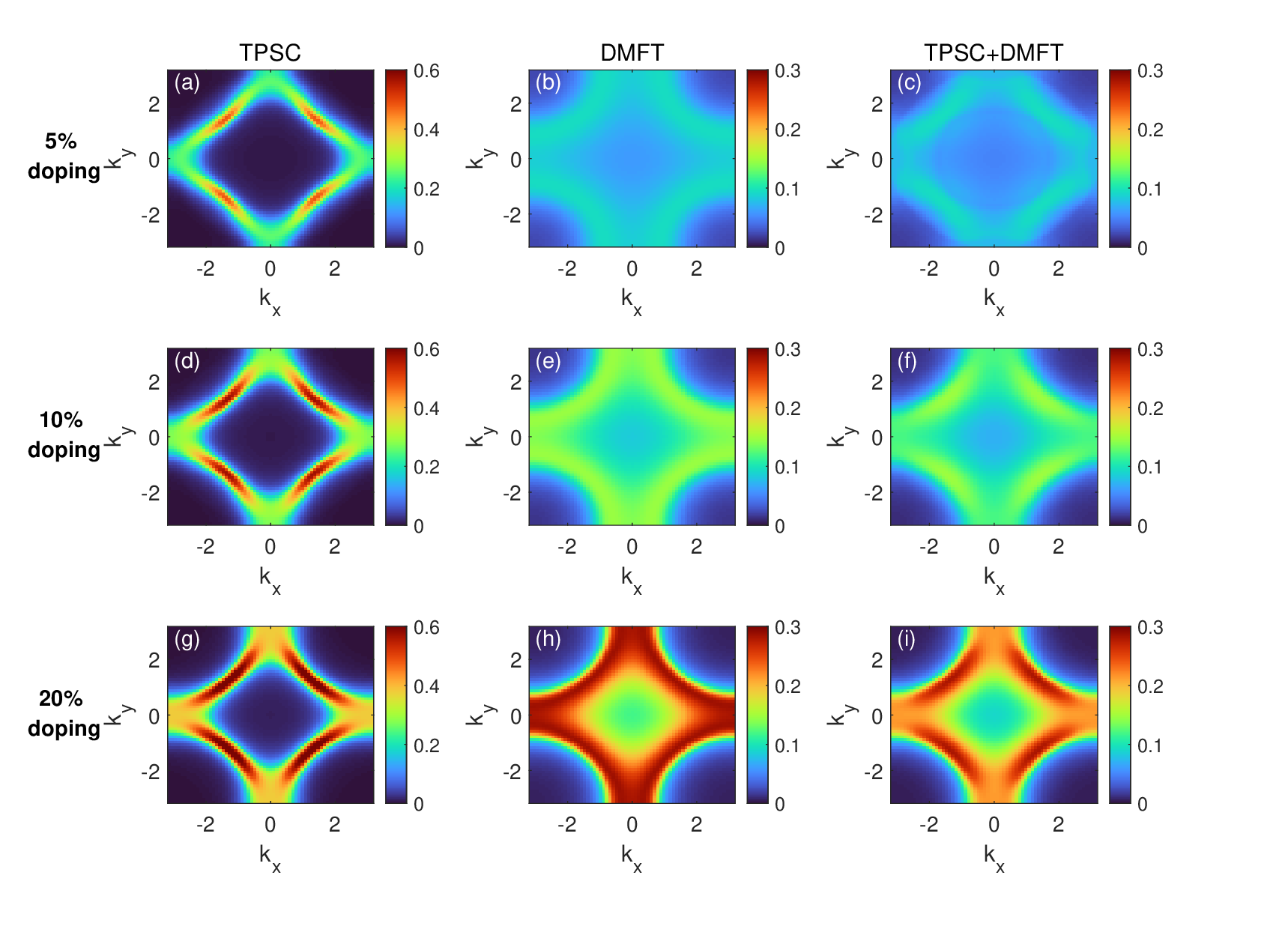}
\caption{Fermi surface of the hole-doped Hubbard model with $t'=-0.3t$ at temperature $T=0.05$. Results obtained using the same method are grouped into columns, while the rows correspond to different doping levels. Panels (a)-(c) represent $5\%$ doping, (d)-(f) $10\%$ doping, and (g)-(i) $20\%$ doping. For each row, panels (a), (d), (g) show results obtained from TPSC, (b), (e), (h) from DMFT, and (c), (f), (i) from TPSC+DMFT.}
\label{fig10}
\end{figure*}

Examining the results within the same row reveals TPSC's inability to generate features such as the (Mott) gap between the quasi-particle band and the upper Hubbard band, which is present in both the DMFT and TPSC+DMFT spectra. While the occupied part of the TPSC spectral function (below the Fermi surface) is similar to the DMFT and TPSC+DMFT counter parts, the quasi-particle band is wider and its energy separation to the lower band is smaller. Overall, the TPSC+DMFT spectra are closer to the DMFT results, although the upper band exhibits unphysical distortions introduced by the TPSC self-energy. These distortions can be mitigated by considering only the nonlocal self-energy from the transverse channel. To better illustrate the effects of the nonlocal corrections on the low-energy electronic structure, we next examine the Fermi surface results.

\subsubsection{Fermi surfaces}

The Fermi surfaces of the doped Mott insulators serve as an excellent benchmark for validating our method, as they can be directly compared with ARPES experiments~\cite{ronning2003evolution,armitage2001anomalous,armitage2002doping}. Previous studies using DMFT extensions that incorporate nonlocal fluctuations, such as cluster DMFT~\cite{civelli2005dynamical,kyung2006pseudogap,sakai2012cluster,sakai2009evolution} and valence-bond DMFT~\cite{ferrero2009pseudogap}, have provided valuable insights, although some of them rely on interpolations of the momentum dependence. These studies revealed the formation of a Fermi pocket centered at the nodal point $\mathbf{k}=(\pi/2,\pi/2)$ in the underdoped regime, which extends into an arc at higher doping levels, with diminishing nonlocal corrections in the overdoped regime. For comparison, we plot the Fermi surfaces obtained using TPSC, DMFT, and TPSC+DMFT at different doping levels in Fig.~\ref{fig10}.

As the hole doping increases, the spectral intensities also increase in all methods, consistent with the trends observed in Fig.~\ref{fig9}. However, while the DMFT results show a uniform intensity along the Fermi surface, the TPSC results display arcs concentrated near the nodal point, due to the large imaginary self-energy components at the antinodal points. The TPSC+DMFT results combine the shorter quasi-particle lifetimes of DMFT (width and intensity of the signal) with the momentum-dependent lifetimes of TPSC. For $5\%$ doping, discontinuities in Figs.~\ref{fig10}(c) suggest that TPSC+DMFT struggles in the underdoped regime, although the pocket-like feature is qualitatively consistent with the expected result. At $10\%$ doping, a clear Fermi arc forms (Fig.~\ref{fig10}(f)), which expands into an almost full (large) Fermi surface at $20\%$ doping (Fig.~\ref{fig10}(i)). While TPSC+DMFT may overestimate nonlocal effects in the overdoped regime, it agrees well with other methods at moderate doping levels. Given the much cheaper cost compared to cluster DMFT and D$\Gamma$A, the high momentum resolution, and the lack of uncertainties associated with analytical continuation, this demonstrates the usefulness of the TPSC+DMFT approach for studying doped Mott insulators.

\begin{figure}[t]
\includegraphics[width=\linewidth]{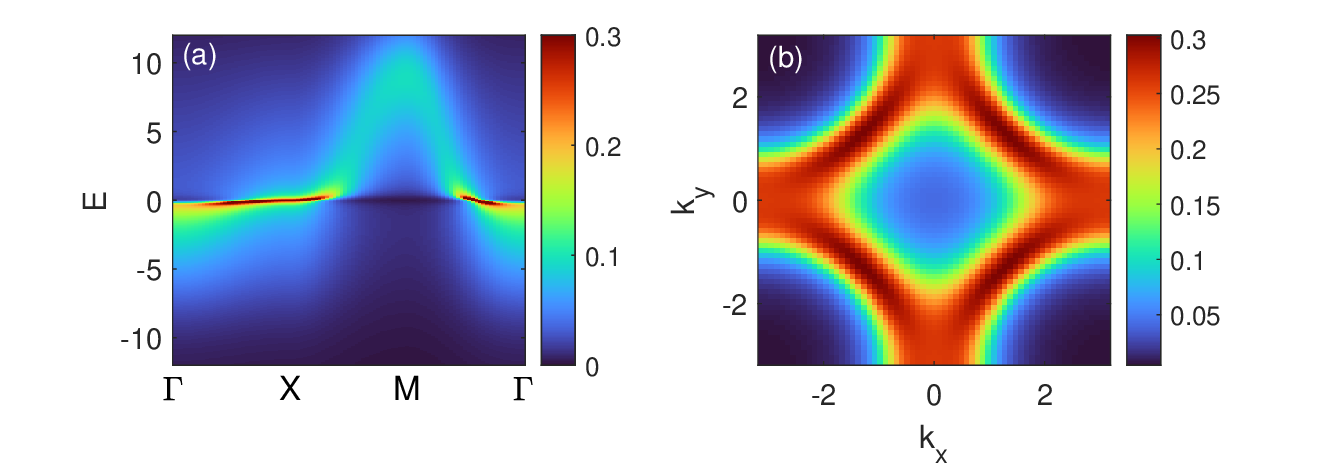}
\caption{FLEX spectral function (a) and Fermi surface (b) for the $20\%$ hole-doped Hubbard model with $t'=-0.3t$ at temperature $T=0.05$.}
\label{fig11}
\end{figure}

As a supplement, we also plot the FLEX spectral function and Fermi surface for the $20\%$ hole-doped system in Fig.~\ref{fig11}. FLEX fails to converge at lower doping levels and produces a poor spectral function (without Hubbard band features) at 20\% doping. In Fig.~\ref{fig11}(b), the Fermi surface exhibits a non-uniform structure, similar to TPSC and TPSC+DMFT. However, the spectral function in Fig.~\ref{fig11}(a) is qualitatively different from the results of the latter methods. The spectral weight is almost entirely concentrated in a narrow band near the Fermi surface and the very blurred structure of the occupied states appears incorrect. This indicates that FLEX is unsuitable for describing doped Mott insulators.

\section{Conclusions}\label{section4}

We used a real-frequency implementation of the TPSC+DMFT method to investigate the correlated electronic structure of the two-dimensional Hubbard model. In the half-filled case with moderate interaction strengths, one-shot TPSC+DMFT captures the expected pseudogap features associated with nonlocal spin fluctuations, as well as the renormalized quasi-particle band and Hubbard band features characteristic of correlated metals. For larger interaction strengths, where the momentum dependence of the self-energy becomes less relevant, the method successfully reproduces Mott insulating behavior.

In the doped Mott regime, TPSC+DMFT provides a qualitatively correct description of the spectral functions and Fermi surfaces. It captures  the formation of Fermi pockets at low doping, which evolve into Fermi arcs and eventually a full Fermi surface at higher doping levels. These features are consistent with theoretical predictions from cluster DMFT calculations~\cite{civelli2005dynamical,Werner2009} and ARPES measurements on cuprates. While TPSC+DMFT may overestimate nonlocal effects in overdoped systems, its performance at moderate doping levels makes it a useful method for studying doped Mott insulators.

More generally, this work demonstrates that TPSC+DMFT is a versatile and computationally efficient approach which combines local and nonlocal correlation effects in Hubbard models. It does not require any interpolations in momentum space, and our real-frequency implementation eliminates the need for analytic continuation. While the current implementation is based on an NCA impurity solver, which leads to an overestimation of correlation effects in the DMFT part~\cite{eckstein2010nonequilibrium}, recently developed higher order steady-state solvers~\cite{Eckstein2024,Kim2024} will overcome this limitation, further enhancing the reliability and range of applicability of the TPSC+DMFT approach.

In principle, the employed real-frequency DMFT and TPSC frameworks would be capable of describing nonequilibrium steady states~\cite{li2021,yan2024spin}, and a potentially interesting application would be the study of spin and charge correlations in such nonequilibrium situations. A stumbling block for such studies is however the lack of self-consistency in the employed TPSC+DMFT scheme. As shown in this paper and discussed in previous works~\cite{simard2023dynamical,yan2024spin}, only the one-shot combination of TPSC and DMFT captures the pseudogap physics, while a self-consistent feedback of the interacting Green's function into the TPSC calculation leads to a physically incorrect solution. While this is a common problem of self-consistent schemes~\cite{gukelberger2015dangers}, it essentially prevents some interesting applications. The description of a photo-doped Mott state with a given nonthermal distribution~\cite{Kuenzel2024,Kim2024}, for example, would require a self-consistent adjustment of the spectral function and occupation.  

\acknowledgements

We thank S. Ray for providing the DMFT solver and for helpful discussions. This work was supported by SNSF Grant No. 200021-196966. The calculations were run on the beo06 cluster at the University of Fribourg.

\appendix
\section{Self-Consistent Variants of TPSC and TPSC+DMFT}
\label{app_A}

Here, we present results from the self-consistent variants of TPSC and TPSC+DMFT, which are TPSC+GG and TPSC+GG+DMFT. As mentioned in Sec.~\ref{section2}, TPSC+GG substitutes the obtained $G^{(2)}$ back into $G^{(1)}$ and iterates the TPSC procedure until self-consistency. TPSC+GG+DMFT is an analogous scheme, where $G^{(2)}$ is obtained from DMFT with nonlocal corrections from TPSC.

\begin{figure}[t]
\includegraphics[width=\linewidth]{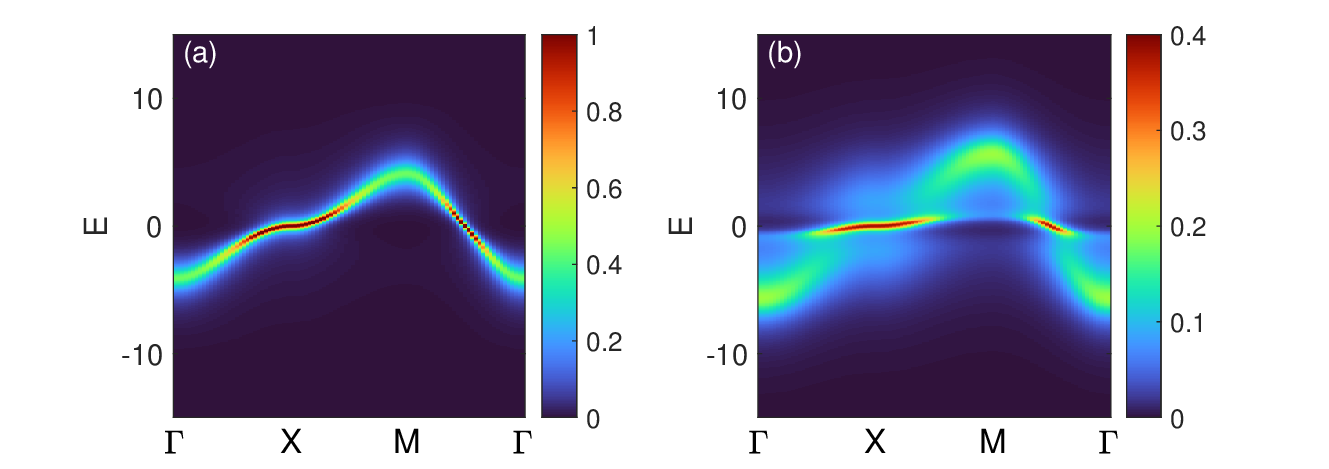}
\caption{Spectral function for the half-filled Hubbard model with $t'=0$ at temperature $T=0.25$ calculated by TPSC+GG (a) and TPSC+GG+DMFT (b).}
\label{figA1}
\end{figure}

\begin{figure}[t]
\includegraphics[width=\linewidth]{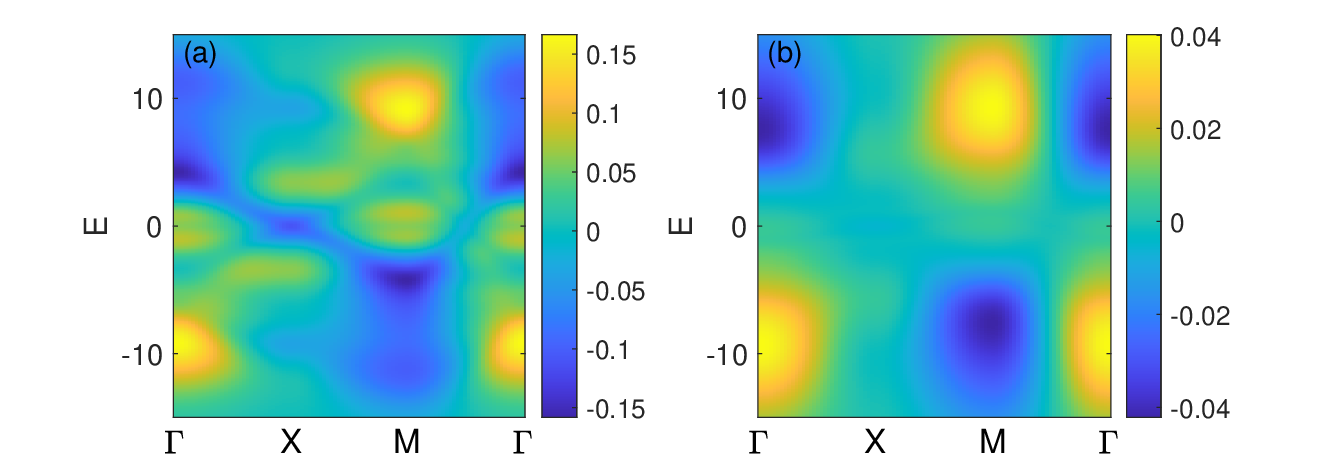}
\caption{Imaginary part of the nonlocal self-energy for the half-filled Hubbard model with $t'=0$ at temperature $T=0.25$ calculated by TPSC+GG (a) and TPSC+GG+DMFT (b).}
\label{figA2}
\end{figure}

The spectral functions of these two methods, calculated using the same parameters as in Fig.~\ref{fig3}, are shown in Fig.~\ref{figA1}. Neither method reproduces the pseudogap feature near the Fermi surface. The result from TPSC+GG, shown in Fig.~\ref{figA1}(a), is close to the FLEX result but exhibits a higher peak intensity. On the other hand, the result from TPSC+GG+DMFT, shown in Fig.~\ref{figA1}(b), closely resembles that of DMFT.

To further examine the correlation effects in the results of Fig.~\ref{figA1}, we plot the corresponding nonlocal self-energies in Fig.~\ref{figA2}. Qualitatively, both methods display flipped bands with negative imaginary values. However, the intensities are significantly lower than those in Fig.~\ref{fig4}, so that these self-energies fail to support the formation of a pseudogap near the antinodal point. Moreover, the intensity of the nonlocal self-energy in Fig.~\ref{figA2}(b) is even lower than that in Fig.~\ref{figA2}(a), particularly in the low-energy regime. As a result, the TPSC+GG+DMFT spectra are almost indistinguishable from those of DMFT.

In summary, the self-consistent variants of TPSC and TPSC+DMFT do not perform well in capturing the pseudogap physics. This finding highlights the challenges in achieving self-consistency while maintaining the essential features of nonlocal fluctuations.

%\addbibresource{mybibtex}
%\printbibliography
%\bibliographystyle{abbrv}  % 使用缩写格式
\bibliography{mybibtex}

\end{document}